\documentclass[usenatbib]{basi}
\usepackage[T1]{fontenc}
\usepackage[british]{babel}
\usepackage[varg]{txfonts}
\usepackage{rotating}
\usepackage{dcolumn}
\begin{document}
\title[Wideband mosaic imaging accuracy]{Wideband Mosaic Imaging with the VLA - quantifying faint source imaging accuracy}
\author[U.~Rau et~al.]%
       {U.~Rau$^1$\thanks{email: \texttt{rurvashi@nrao.edu}},
       S.~Bhatnagar$^1$ and F.~N.~Owen$^1$\\
       $^1$National Radio Astronomy Observatory, Socorro, New Mexico, USA}

\pubyear{2014}
\volume{00}
\pagerange{\pageref{firstpage}--\pageref{lastpage}}
%\status{submitted}

\date{Received --- ; accepted ---}

\maketitle
%------------------------------------------------------------------------------%
% abstract and keywords                                                        %
%------------------------------------------------------------------------------%
\label{firstpage}

\begin{abstract}
A large number of deep and wide-field radio interferometric surveys are being designed
to measure accurate statistics of faint source populations. Most require mosaic observations, 
and expect to benefit from the sensitivity provided by broad-band instruments. 
In this paper, we present preliminary results from a comparison of several wideband imaging
methods in the context of how accurately they reconstruct the intensities and 
spectral indices of micro-Jy level sources. 
\end{abstract}


%------------------------------------------------------------------------------%
% main text of the paper, using \section, \subsection, \subsubsection          %
%------------------------------------------------------------------------------%
\section{Introduction}\label{s:intro}

The recent upgrade of the Very Large Array (VLA) has resulted in a greatly increased
imaging sensitivity due to the availability of large instantaneous bandwidths at the
receivers and correlator. A considerable amount of telescope time has been allotted for
large survey projects that need deep and sometimes high dynamic range imaging 
over fields of view that span one or more primary beams. 
%Data sizes range from a few hundred 
%Gigabytes up to a few Terabytes and contain a large number of frequency channels 
%for one or more pointings.
In this imaging regime, traditional algorithms have limits in the achievable dynamic range
and accuracy with which weak sources are reconstructed. 
Narrow-band approximations of the sky brightness 
and instrumental effects result in sub-optimal continuum sensitivity
and angular resolution. Narrow-field approximations that ignore the time-, frequency-, and
polarization dependence of antenna primary beams prevent accurate reconstructions 
over fields of view larger than the inner part of the primary beam. Mosaics constructed by
stitching together images reconstructed separately from each pointing often have a lower
imaging fidelity than a joint reconstruction. Despite these drawbacks, all these methods 
are easy to apply using readily available and stable software and are therefore 
used regularly. 

More recently-developed algorithms that address the above shortcomings also exist.
Wide-band imaging algorithms \citep{MFCLEAN, MSMFS} make use of the
combined multi-frequency spatial frequency coverage while reconstructing both the
sky intensity and spectrum at the same time. Wide-field imaging algorithms
\citep{WPROJ, APROJ}
include corrections for instrumental effects such as the w-term and antenna 
aperture illumination functions. Wideband A-Projection \citep{WBAWP}, a 
combination of the two methods mentioned above 
separates the frequency dependence of the sky from that of the
instrument during wideband imaging. Finally, an algorithm to perform a joint mosaic reconstruction
\citep{OLDMOSAIC}  along with a wideband sky model and wideband primary beam correction 
has recently been demonstrated to work accurately and is currently being commissioned
\citep{WBMOS}(in prep).
These methods provide superior numerical results compared to traditional methods
but they require all the data to be treated together during the reconstruction and 
need specialized software implementations that are optimized for the large amount
of data transport and memory usage involved in each imaging run.

With so many methods to choose from and various trade-offs between numerical 
accuracy, computational complexity and ease of use, it becomes important to identify
the most appropriate approach for a given imaging goal and to quantify the errors
that would occur if other methods are used.
This paper describes some preliminary results based on a series of simulated tests
of deep wide-band and wide-field mosaic observations with the VLA. 

\section{Data Simulation}
A sky model was chosen to contain a set of 8000 point sources spanning
one square degree in area. Intensities ranged between $1 \mu Jy$
and $7 mJy$ plus one bright $100 mJy$ source, and followed a realistic source
count distribution.  Spectral indices ranged between 0.0 and -0.8 with a peak in the
spectral index distribution at -0.7 plus a roughly Gaussian distribution
around -0.3 with a width of 0.5. 
This source list is a
subset of that available from the SKADS/SCubed simulated sky project \citep{SKADS}.

Observations were simulated for a VLA mosaic at D-config and C-band with 46 pointings
(of primary beams 6 arcmin in HPBW at 6 GHz) spaced 5 arcmin apart. 
16 channels (or spectral windows) were chosen to span the frequency range of 4-8 GHz,
and the $uv$-coverage corresponds to one pointing snapshot every 6 minutes, tracing
the entire mosaic twice within 8.8 hours. 

%Two types of datasets were simulated. One was for a VLA single pointing at C-config and L-band
%with 16 channels (or spectral windows) between 1 and 2 GHz. The $uv$-coverage was
%a series of snapshots 20 minutes apart, for 4 hours. This is a relatively sparse coverage
%and the results were later compared with a dataset with snapshots taken every 2 minutes 
%instead. The HPBW of the primary beam at L-band is 30arcmin and therefore covers the
%central part of the simulated sources. 
%The second dataset was for a VLA mosaic at D-config and C-band with 46 pointings (of primary
%beams 6arcmin in HPBW) to cover roughly the same patch of sky at a comparable angular
%resolution.  

Visibilities were simulated per pointing, using the WB-A-Projection de-gridder \citep{WBAWP} 
which used complex antenna aperture illumination functions to model primary beams that rotate
with time, scale with frequency, and have polarization squint. No noise was added.

\section{Imaging Algorithms}

\begin{figure}
 \includegraphics[width=5.5in]{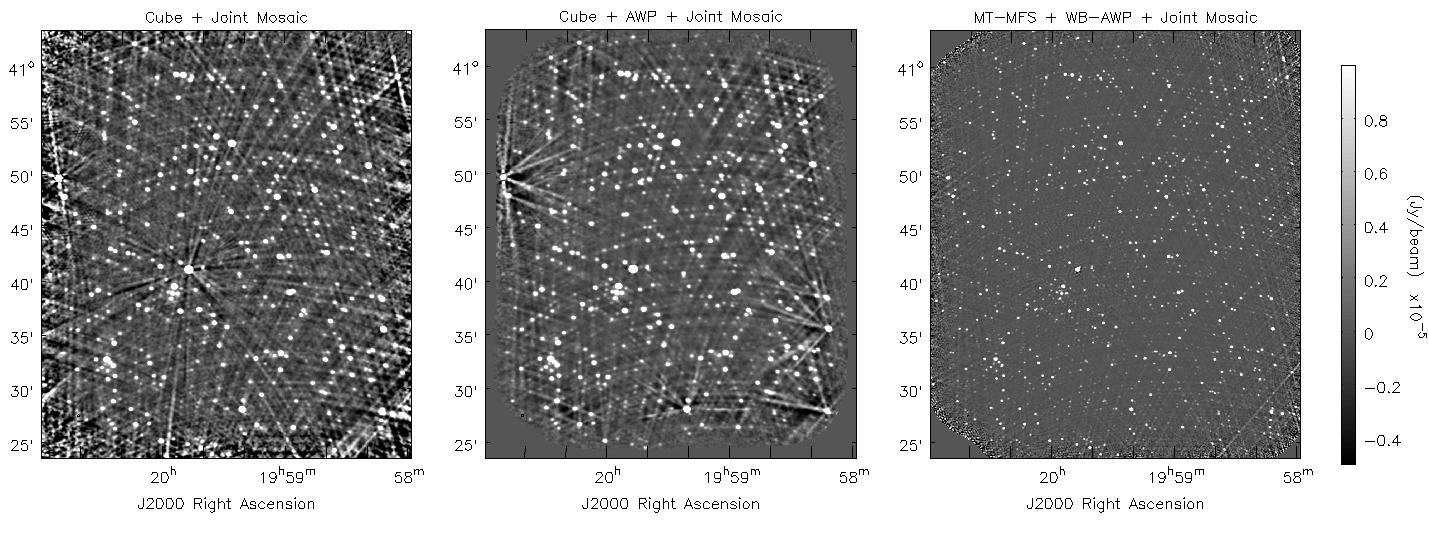}
\caption{Restored continuum intensity images : (LEFT) Cube + Joint Mosaic, 
(MIDDLE) Cube + AWP + Joint Mosaic, (RIGHT) MT-MFS + WB-AWP + Joint Mosaic.
Intensity range shown : $-5\mu Jy$ to $+10\mu Jy$. }
\label{fig.images}
\end{figure}

The wideband mosaic dataset described above was imaged in a variety of ways. In all cases the data
products were continuum intensity images and spectral index maps.
Figure.~\ref{fig.images} shows restored continuum mosaic images from 
the first, third and fifth algorithms described below. 

%\paragraph{Wideband Single Pointing}
%\begin{enumerate}
%\item MT-MFS : Multi-term multi-frequency synthesis that solves for intensity and spectrum
%at the same time and combines the sky spectrum and the frequency dependence of 
%the primary beam, followed by a 
%post-deconvolution correction of an average primary beam and its spectrum. 
%\item MT-MFS + WB-AWP : Multi-frequency synthesis with a wideband sky model for
%only the sky spectrum, along with wideband A-Projection that eliminates the frequency
%dependence of the primary beam before the modeling process. A post-deconvolution
%correction of only the average primary beam is required in this case.
%\item Cube : Each channel (spectral window) is reconstructed independently with 
%separate post-deconvolution primary-beam corrections. All images are them smoothed
%to the angular resolution of the lowest frequency in the observation before fitting 
%spectral models and collapsing channels to form a continuum image.
%\item Cube + AWP : Same as Cube, but with narrow-band A-Projection per channel
%to account for beam rotation and squint. 
%\end{enumerate}
%
%\paragraph{Wideband Mosaic}
\begin{enumerate}
\item Cube + Joint Mosaic : 
A joint mosaic reconstruction is performed separately per
channel (spectral window). A mosaic primary beam correction is then done per channel, and
all channels are smoothed to the angular resolution at the lowest frequency in the observation.
Spectral models are then fit for each pixel/source and a weighted frequency average is done to form
a continuum image.
\item Cube + AWP + Joint Mosaic : Same as above, but with narrow-band
A-Projection per channel (spectral window) to account for beam rotation and squint.
\item MT-MFS + Stitched Mosaic : Each pointing is imaged separately using
multi-term multi-frequency-synthesis followed by post-deconvolution wideband primary beam 
correction and adding all image patches together by weighting them by an 
average primary beam. 
\item MT-MFS + WB-AWP + Stitched Mosaic: Each pointing is imaged
separately using MT-MFS along with wideband A-Projection that eliminates the frequency
dependence of the primary beam before the image modeling process. A post deconvolution
correction of only the intensity image is required before a weighted average is done to
combine images from all pointings.
\item MT-MFS + WB-AWP + Joint Mosaic : All data are imaged together, with a 
wideband sky model and antenna-, time-, frequency- and polarization-dependent 
primary beam correction during imaging with wideband A-Projection.
\end{enumerate}

\section{Tests and Results}

The intensity and spectral index maps produced by the above algorithms were
compared with the known simulated sky.  For each output image, the simulated sky 
model image was first smoothed to match its angular resolution, and then pixel values
were read off from both images at all the locations of the true source pixels. 
Histograms were plotted for $I/I_{true}$ where deviations from 1.0 indicate
relative flux errors and for $\alpha - \alpha_{true}$ where deviations from 0.0 indicate
relative errors in spectral index. 
All histograms were made with multiple intensity ranges 
and over different fields of view to look for trends in the errors. 
%
%The ratio
%$I/I_{true}$ was plotted as a histogram where a spread around 1.0 indicates relative flux errors.
%For spectral index, $\alpha - \alpha_{true}$ was 
%plotted as a histogram where a spread around 0.0 indicates relative errors
% in spectral index.. All histograms were made with different
%intensity ranges and over different fields of view to look for trends in the errors. 
%
The mean and half-width of each of the resulting distributions (over different intensity
ranges) for the first, third and fifth method described above  
are listed in Table~\ref{tab:errors}.  
%The first three columns represent $I/I_{true}$ and the last two columns 
%represent $\alpha - \alpha_{true}$. 
Spectral index reconstructions for the weakest
sources $<5\mu Jy$ were not included as all the methods were inaccurate.
These numbers show that cube methods have wider distributions for all intensity ranges.
This is primarily because of weak sources that are not detected in single channel images but 
appear as confused undeconvolved sources in the continuum image. 
The achieved mean values in both intensity and spectral index
show that accurate handling of the primary beam (via A-Projection)
is required in order to recover the intensity and spectral index to within a few percent,
particularly for weak sources. 
These results are part of a larger study (to be described in an upcoming publication)
that includes single pointing tests to evaluate effects
%In addition to these tests, a series of single pointing tests were done to evaluate effects
of sparse $uv$-coverage, the use of masks during deconvolution, the accuracy of
beam polarization correction, a method to deal with undeconvolved sources in cube
methods, sources not on pixel centers, baseline based averaging, visibility noise, and
various choices and numerical approximations within the reconstruction algorithms.

\begin{table}
  \caption{Intensity and Spectral Index reconstruction accuracy as a function of source intensity}\label{tab:errors}
  \medskip
  \begin{center}
    \begin{tabular}{cccccc}\hline
      Method  & $I/I_{true}$  & $I/I_{true}$ &  $I/I_{true}$ & $\alpha - \alpha_{true}$ & $\alpha - \alpha_{true}$\\
          Intensity Range          & $>20\mu Jy$  & $5 - 20 \mu Jy$ & $ < 5\mu Jy$ & $>50\mu Jy$  & $10 - 50 \mu Jy$ \\
      \hline
        Cube   & 0.9 $\pm$ 0.1 &  0.9 $\pm$ 0.3 & 0.9 $\pm$ 0.5 & -0.5 $\pm$ 0.2 &  -0.6 $\pm$ 0.5 \\
        Cube + AWP  & 1.0 $\pm$ 0.05 &  1.0 $\pm$ 0.2 & 1.0 $\pm$ 0.3 & -0.15 $\pm$ 0.1 &  -0.1 $\pm$ 0.25  \\
        MTMFS + WB-AWP & 1.0 $\pm$ 0.02 &  1.0 $\pm$ 0.04 & 1.0 $\pm$ 0.15  & -0.05 $\pm$ 0.05 &  -0.1 $\pm$ 0.2\\
        \hline
    \end{tabular}\\[5pt]
%    \begin{minipage}{7cm}
%      \small For each method, the reconstruction accuracy is quantified separately for different 
%intensity ranges, as the mean and half-width of the distribution of values about the true value.
%The first three columns represent $I/I_{true}$, where deviations from 1.0 indicate relative flux
%errors. The last two columns represent $\alpha - \alpha_{true}$ where deviations from 0.0 
%indicate relative errors in spectral index. 
%    \end{minipage}
  \end{center}
\end{table}

%------------------------------------------------------------------------------%
%\section*{Acknowledgements}
%
%We would like to thank the NRAO for the use of their CASA software libraries for this work.

%------------------------------------------------------------------------------%
% bibliography: produced from ADS using custom format of                       %
%                                                                              %
%     %z132 \\bibitem[%\2%(y)%\3m]%{R}\n   %\8.1g,%\Y,%\q,%\V,%\ p             %
%------------------------------------------------------------------------------%

\label{lastpage}
%------------------------------------------------------------------------------%

\begin{thebibliography}{}

\bibitem[Sault et al.(1994)Sault \& Wieringa]{MFCLEAN}
  Sault,R.J. \& Wieringa,M., 1994, A\&A Suppl.Ser, 108, 585

\bibitem[Rau et al.(2011)Rau, \& Cornwell]{MSMFS}
  Rau, U. \& Cornwell, T.~J., 2011, A\&A, 532, A71

\bibitem[Cornwell et al.(2008)Cornwell et.al.]{WPROJ}
  Cornwell, T.~J., Golap, K., Bhatnagar, S., 2008, IEEE Sel. Top. in Sig. Proc., vol2, 647-657

\bibitem[Bhatnagar et al.(2008)Bhatnagar et al]{APROJ}
  Bhatnagar, S., Cornwell, T.~J., Golap, K., Uson, J.~M., 2008, A\&A, vol 487, 419-429

\bibitem[Bhatnagar et al.(2013)Bhatnagar et al.]{WBAWP}
  Bhatnagar, S., Rau, U, Golap, K., ApJ, 770, 91

\bibitem[Cornwell, T.J.(1998)Cornwell]{OLDMOSAIC}
  Cornwell, T.~J., A\&A, 202,316

\bibitem[Rau et al.(2014)Rau et al]{WBMOS}
  Rau, U., Bhatnagar, S., Golap, K., (in prep)

\bibitem[Wilman et al.(2008)Wilman et al]{SKADS}
	Wilman, R. ~J., Miller,L., Jarvis, M. ~J., Mauch, T., Levrier, F., Abdalla, F. ~B., Rawlings, S., Klöckner, H.~R., Obreschkow, D., Olteanu, D., Young, S, MNRAS, Vol 388, 1335-1348

\end{thebibliography}
\end{document}